\documentstyle[epsfig,epsf]{article}
\def\Journal#1#2#3#4{{#1}{\bf #2}, #3 (#4)}
\def\nmne {\nu_\mu \longleftrightarrow \nu_e}
\def\nmnt {\nu_\mu \longleftrightarrow \nu_\tau}
\def\nmns {\nu_\mu \longleftrightarrow \nu_{sterile}}
\def\ne {\nu_e}
\def\nm {\nu_\mu}
\def\nt {\nu_\tau}
\def\Dm {\Delta m^2}
\def\s2t {\sin^2 2\theta}

\def\NIMA{{\em Nucl. Instr. Meth.} {\bf A}}
\def\PLB{{\em Phys. Lett.} {\bf B}}
\def\PRL{\em Phys. Rev. Lett.~}
\def\PRD{{\em Phys. Rev.} {\bf D}}
\def\ZPC{{\em Z. Phys.} {\bf C}}
\def\NPBPS{{\em Nucl. Phys.} B {\em Proc. Suppl.}~}
\def\EPJC{{\em Eur. Phys. J.} {\bf C}}
\def\AP{\em Astrop. Phys.~}
\def\PAN{\em Phys. Atom. Nucl.~}
\def\BJP{\em Braz. J. Phys.~}
\def\APPB{{\em Acta Phys. Polon.} {\bf B}}

\begin{document}

\begin{center}
{\Large {\bf ATMOSPHERIC NEUTRINO OSCILLATIONS}}
\end{center}

\vskip .7 cm

\begin{center}
G. GIACOMELLI and M. GIORGINI \par~\par
{\it Dept of Physics, Univ. of Bologna and INFN, \\
V.le C. Berti Pichat 6/2, Bologna, I-40127, Italy\\} 

E-mail: giacomelli@bo.infn.it , giorginim@bo.infn.it

\par~\par

Lectures at the 7$^{th}$ School on Non-Accelerator Astroparticle Physics,  
\\ Trieste, Italy, 26 July - 6 August 2004. 

\vskip .7 cm
{\large \bf Abstract}\par
\end{center}

{\normalsize 
The latest results from the Soudan 2, MACRO and 
SuperKamiokande experiments on atmospheric neutrino oscillations are
summarised and discussed. In particular a discussion is made on the Monte 
Carlo simulations used for the atmospheric neutrino flux.}

\vspace{5mm}

\section{Introduction}\label{sec:intro}
Atmospheric neutrinos are generated in the decays of hadrons produced in high
energy cosmic ray (CR) interactions. A high energy primary cosmic ray, proton 
or nucleus, interacts 
in the upper atmosphere producing  a large number of pions 
and kaons, which decay yielding  muons and muon
neutrinos; also the muons decay yielding muon and electron neutrinos.
 The ratio of the
numbers of muon to electron neutrinos is about 2 
and $N_{\nu}/N_{\overline\nu} \simeq 1$.
Atmospheric  neutrinos are produced in a spherical
surface at about 10-20 km above ground and they proceed towards the earth. 

If neutrinos have non-zero masses, one has to consider 
the {\it weak flavour eigenstates} $~\ne,~\nm,~\nt$ and the {\it mass
eigenstates} $~\nu_1,~\nu_2,~\nu_3$. 
The weak flavour eigenstates $\nu_l$ are linear combinations of the mass 
eigenstates $\nu_m$ through the elements of the mixing matrix $U_{lm}$:

\begin{equation}
\nu_l = \sum_{m=1}^3 U_{lm}\ \nu_m
\end{equation}

For 2 flavour eigenstates ($\nm,~\nt$) and
2 mass eigenstates $(\nu_2,~\nu_3)$ 

\begin{equation}
\left\{ \begin{array}{ll}
      \nm =~\nu_2 \cos\ \theta_{23} + \nu_3 \sin\ \theta_{23} \\
      \nt=-\nu_2\sin\ \theta_{23} + \nu_3\cos\ \theta_{23} 
\end{array} 
\right. 
\end{equation} 
\noindent where $\theta_{23}$ is the mixing angle. The 
survival probability of a $\nm$ beam is
\begin{equation}
P(\nm \rightarrow \nm) = 1- \sin^2 2\theta_{23}~\sin^2 \left( 
{\frac{1.27 \Dm \cdot L} {E_\nu}} \right)
\end{equation}
\noindent where $\Dm=m^2_3-m^2_2$, $L$ is the distance travelled by the 
$\nu$ from production to detection. The probability 
for the initial $\nm$ to oscillate into a $\nt$ is
\begin{equation}
P(\nm \rightarrow \nt) = 1 - P(\nm \rightarrow \nm) =
 \sin^2 2\theta_{23}~\sin^2 \left( { 
{1.27 \Dm \cdot L}\over {E_\nu}} \right)
\label{eq:prob_osc}
\end{equation}
Atmospheric neutrinos are well suited for the study
of neutrino oscillations, since they have energies from a fraction of GeV up 
to more than 100 GeV and they travel distances $L$ from few tens of km 
up to 13000 km; thus $L/E_\nu$ ranges from $\sim 1$ km/GeV to $10^5$ 
km/GeV. Moreover one may consider that there are two identical sources for
a single detector: a near one (downgoing neutrinos) and a far one (upgoing
neutrinos). Atmospheric neutrinos are particularly useful to study oscillations
for small $\Dm$, and matter effects can be studied with their high
energy component. 

The early water Cherenkov detectors and the tracking calorimeters
measured $\nm$ 
and $\ne$ charged current interactions. The results were
expressed in terms of the double ratio $R^\prime =R_{obs}/R_{MC}$, where
$R_{obs} = (N_{\nm} / N_{\ne})_{obs}$ is the ratio of observed 
$\mu$ and $e$ events and $R_{MC} = (N_{\nm}/ N_{\ne})_{MC}$ 
is the same ratio for Monte Carlo (MC) events. The $R^\prime$ double 
ratios from IMB \cite{imb} and 
Kamiokande \cite{kamioka} were smaller than expectations, while the
NUSEX \cite{nusex} and Frejus \cite{frejus} $R^\prime$ agreed with 
expectations. The Baksan \cite{baksan} scintillation telescope detected 
upthroughgoing muons at the expected rate but gave indications of 
 an anomalous angular distribution. Later, the Soudan 2 tracking and 
shower calorimeter detector confirmed the anomaly in the $\nm/\ne$ double 
ratio for contained events \cite{soudan}. MACRO reported 
in 1995 a measurement of upthroughgoing muons coming from $\nm$ of 
 $\langle E_\nu \rangle \sim 50$ GeV, in which there was an 
anomalous zenith distribution and a deficit in the total number of 
observed upgoing muons \cite{macro-95}. SuperKamiokande (SK) confirmed the 
anomalous double ratio
 and provided a wealth of informations for sub-GeV and multi-GeV 
$\nm$, $\ne$ and for higher energy upthroughgoing muons and stopping 
muons. In 1998 Soudan 2, MACRO and SK provided strong indications 
in favour of $\nmnt$ oscillations [8-11]. 
 After 1998 new results were presented by the 3 
experiments [12-14]. Here we shall review their results. 

\section{Atmospheric neutrino flux calculations}\label{sec:MC}
In the past use was made of unidimensional Monte Carlo 
codes, Bartol96 \cite{Bartol96} and HKKM95 \cite{honda96}. Recently new 
improved MC predictions for neutrino fluxes were made available 
by the HKKM01 \cite{honda01} and FLUKA \cite{FLUKA} groups. 
They include three dimensional calculations of hadron production and decays
and of neutrino interactions, improved hadronic model and new fits of the
primary CR flux. The two MCs yield predictions for the non oscillated 
and oscillated $\nm$ fluxes equal to within few \%. The shapes of the
angular distributions for oscillated and non oscillated 
Bartol96, HKKM95, new FLUKA
and new HKKM01 fluxes are the same to within few \%.
The absolute values of the MACRO upthroughgoing muon data are about $25\%$
higher than those predicted by the new FLUKA and HKKM01 
MC \cite{macro-nu}, while the shapes of the oscillated and non oscillated
angular distributions 
differ by no more than 5\%, see Sec. \ref{sec:MACRO}.

\begin{figure}
\begin{center}
\mbox{\epsfig{figure=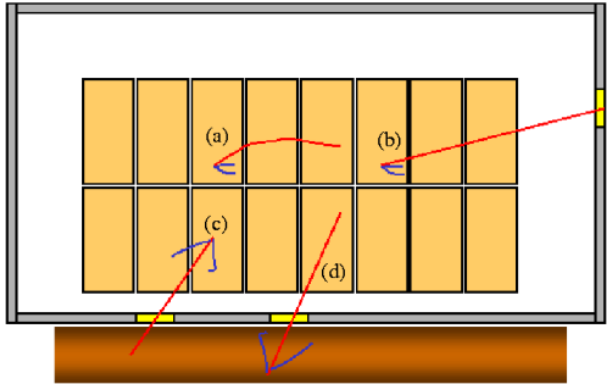,height=5.1cm}}
\caption {Longitudinal cross section of the Soudan 2 detector and observed 
event topologies: (a) Fully Contained Events 
$\langle E_\nu \rangle \sim 1$ GeV, (b) Partially Contained Events
$\langle E_\nu \rangle \sim 6$ GeV, (c) In-down muons 
$\langle E_\nu \rangle \sim 2.4$ GeV, (d) Up-stopping muons
$\langle E_\nu \rangle \sim 6.2$ GeV.}
\label{fig:soudan2_detector}
\end{center}
\end{figure}

 A similar situation is found in the SK data \cite{skam-new}, see Sec.
\ref{sec:SK}.
 The electron-like events were in agreement with the HKKM95 \cite{honda96} 
MC predictions in absence of oscillations, while they are higher than the 
HKKM01 \cite{honda01} non oscillated MC, Fig. \ref{fig:sk_zenith}. For the 
muon-like events, the new 
MC predictions are low for the SK data, especially for the high 
energy upthroughgoing events \cite{skam-new}. Previous comparisons between 
the SK muon data and the 
HKKM95 \cite{honda96} predictions showed a global deficit of events and a
zenith distribution in agreement with $\nmnt$ oscillations \cite{skam-new}.

The difference between the new and old MC predictions is very
probably due to the use of a new fit of the cosmic ray data \cite{gaisser}.

 Recent results by the L3C and BESS experiments \cite{L3C-BESS} on the primary 
cosmic ray fit show good agreement with the Bartol96 and HKKM95 
predictions and a disagreement with the new fit of the cosmic 
ray data \cite{gaisser}. 

The calculations of HKKM01 and FLUKA are in good agreement when
using the old fit or the new fit to the primary CR flux; this confirms the 
improvement in the hadronic model.

In SK the evidence for neutrino oscillations lies in the shapes of the angular
distribution and in the ratio of $\mu/e$ data. In MACRO the evidence  
is due mainly to the shape of the high energy angular 
distribution and the prediction is the same in all simulations. 

The MACRO data suggest that the FLUKA normalisation should be raised by 
$\sim 25 \%$ at $E_\nu \sim 50$ GeV and by $12 \%$ at 
$E_\nu \sim 2-3$ GeV \cite{macro-nu}. Similar conclusions are reached by 
SK \cite{skam-new}.

\begin{figure}
 \begin{center}
\mbox{\epsfig{figure=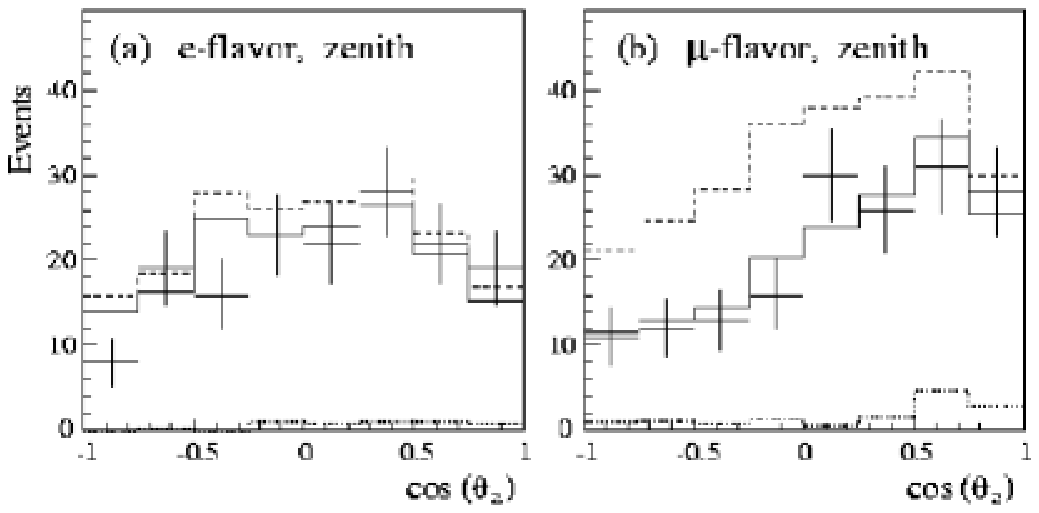,height=4.5cm}}
        \hspace{2mm}
\mbox{\epsfig{figure=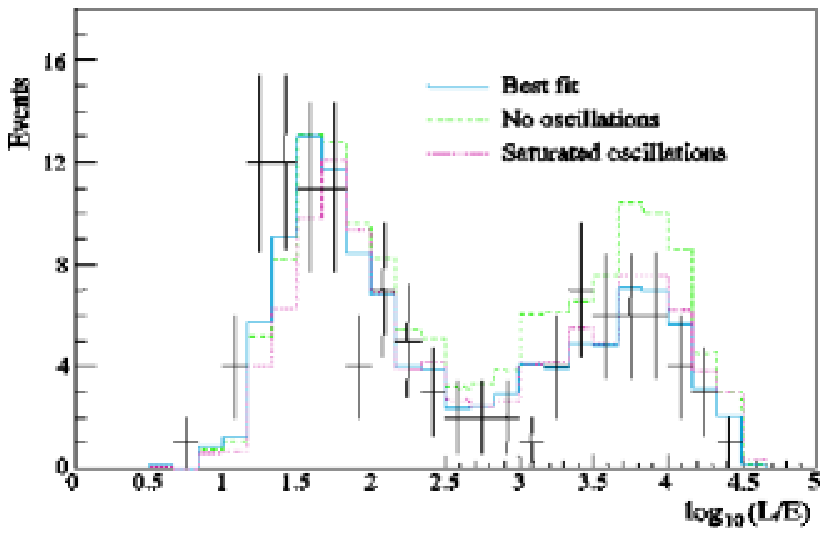,height=4.4cm}}
{\footnotesize  (a) \hspace{4cm} (b) \hspace{4cm} (c)}
\caption{Zenith angle distribution for Hi-Res $e-$flavour (a) and 
$\mu-$flavour (b) events. The points with error bars are the data, the 
dashed and solid histograms are the predicted non oscillated and oscillated 
$\nu$ distributions, respectively, the dotted histograms are the 
contribution of the rock background. (c) Distribution in $\log L/E_\nu$ 
for the Hi-Res Fully Contained events (black crosses) compared with the MC 
predictions for oscillations (solid histogram), no oscillations (dashed 
histogram) and saturated oscillations (dotted 
histogram).}
\label{fig:soudan2_zenith}
\end{center}
\end{figure}

\section{Results from the Soudan 2 experiment}\label{sec:soudan2}
The Soudan 2 experiment used a modular fine grained tracking and showering
calorimeter of 963 t, located 2100 m.w.e. underground in the Soudan Gold 
mine in Minnesota. The
bulk of the mass consisted of 1.6 mm thick corrugated  steel sheets 
 interleaved with drift tubes. The detector was surrounded by an 
anticoincidence shield. Figure \ref{fig:soudan2_detector} shows 
a longitudinal cross section of the apparatus and the topologies of the 
events observed. The final analysis used the Fully Contained High-Resolution 
events \cite{soudan-new}, Table \ref{tab:hi-res}. 

\begin{table}[ht]
\begin{center}
\begin{tabular}{@{}ccc@{}}
\hline
{} &{} &{}\\[-1.5ex]
{} & Data & MC$_{\mbox{no~osc}}$ \cite{Bartol96}\\[1ex]
\hline
{} &{} &{}\\[-1.5ex]
Track & $101.9 \pm 12.7$ & 193.1\\[1ex]
Showers & $146.7 \pm 12.5$ & 179.0\\[1ex]
\hline
\end{tabular}\label{tab:hi-res} 
\caption{Soudan 2 Hi-Res data.}
\end{center}
\end{table}

The fully contained events consist mostly of quasi-elastic 
neutrino reactions, but include a
background of photons and neutrons from cosmic ray muon
interactions in the surrounding rock. 
 The track and shower events for a 5.9 kt$\cdot$yr exposure are summarised in
Table \ref{tab:hi-res}, where they are compared with MC predictions
based on the  Bartol96 neutrino flux \cite{Bartol96}.

After corrections for background and selecting a high resolution (Hi-Res) 
sample of events, the Soudan 2 
double ratio for the whole zenith angle range ($-1 \leq \cos \Theta \leq 1$) 
is $R^\prime=(N_\mu/ N_e)_{DATA}/ (N_\mu/ N_e)_{MC} = 0.69 \pm 
0.12$, consistent with muon neutrino oscillations.

The $\ne$ data agree with the no oscillation MC 
predictions, while the $\nm$ data are lower, except in the forward
zenith bin. The 
double peak structure arises from the acceptance of the apparatus. The 
roughly interpolated 90\% C.L. allowed region in the $\s2t - \Dm$ 
plane, computed using the Feldman-Cousins method \cite{feldman-cousins} 
is shown in Fig. \ref{fig:sk_le_contour}, where it is compared with the allowed
regions obtained by the SK and MACRO experiments.

\section{Results from the MACRO experiment}\label{sec:MACRO}
MACRO was a large area multipurpose underground detector designed
to search for rare events and rare phenomena in the penetrating cosmic
radiation. It was located in Hall B of the Gran Sasso Lab at
an average rock overburden of 3700 m.w.e.; it started data taking with 
part of the apparatus in 1989; it was completed in 1995 and  was running in 
its final configuration until the end of 2000. 
The detector had global dimensions of $76.6\times 12 \times 9.3$ m$^3$;
 vertically it was divided into a lower part,
which contained 10 horizontal layers of streamer tubes, 7 of rock absorbers
and 2 layers of liquid  scintillators, and an upper part which contained the
electronics and was  covered by 1 scintillator layer and 4 layers of
streamer tubes. The sides were covered with 1 vertical scintillator layer 
and 6 of limited streamer tubes \cite{macro-technical}.

\begin{figure}
 \begin{center}
\mbox{\epsfig{figure=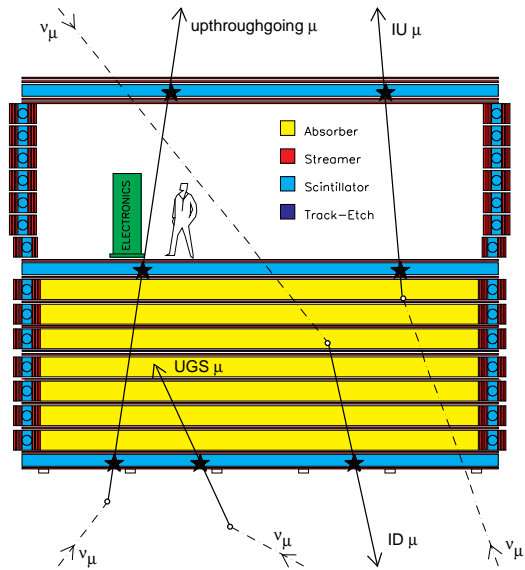,height=7.5cm}}
\caption {Cross section of the MACRO detector. Event topologies induced
by $\nm$ interactions in or around the detector. IU$_\mu=$ semicontained 
Internal Upgoing $\mu$; ID$_\mu=$ Internal Downgoing $\mu$; UGS$_\mu=$ 
Upgoing Stopping $\mu$; Upthroughgoing = upward throughgoing $\mu$.}
\label{fig:macro_topo}
\end{center}
\end{figure}

\begin{figure}
 \begin{center}
\mbox{\epsfig{figure=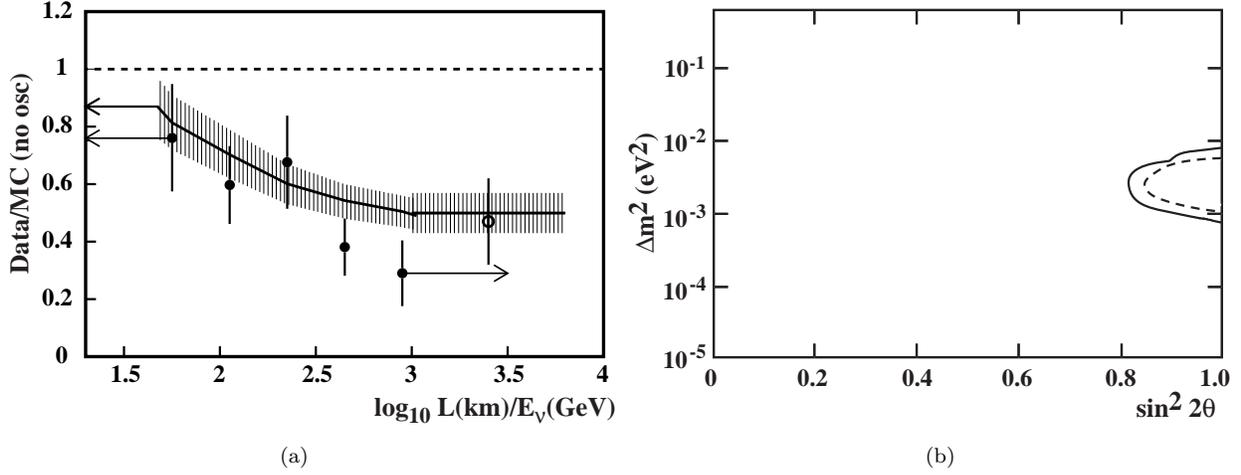,height=5.8cm}}
{\footnotesize  (a) \hspace {8cm} (b)}
\caption {(a) Ratio Data/MC$_{\mbox{no~osc}}$ 
as a function of the estimated $L/ E_\nu$ for the upthroughgoing muon 
sample (black points). The solid line is the MC expectation assuming 
$\Dm = 2.3 \cdot 10^{-3}$ eV$^2$ and $\s2t = 1$. The last point 
(empty circle) is obtained from the IU sample. (b) Interpolated 90\% C.L.
contour plots of the allowed regions in the $\s2t -\Dm $ plane for the 
MACRO data using only the ratios $R_1,R_2,~R_3$ (continuous line) and adding
also the information on the absolute values $R_4,~R_5$ (dotted line).}
\label{fig:le-contour}
\end{center}
\end{figure}

MACRO detected upgoing $\nm$'s via charged current interactions 
$\nm \rightarrow \mu$; upgoing muons were identified with the
streamer tube system (for tracking) and the scintillator system (for
time-of-flight measurement).
The events measured and expected for the 3 measured 
topologies, Table \ref{tab:macro}, and the $L/E_\nu$ distribution, Fig.
\ref{fig:le-contour}a, deviate from MC expectations without 
oscillations; the deviations point to the same $\nmnt$  oscillation 
scenario \cite{macro-nu}.

\begin{table}[ht]
\begin{center}
\begin{tabular}{@{}cccc@{}}
\hline
{} &{} &{} &{}\\[-1.5ex]
{} & Events & MC$_{\mbox{no~osc}}$ \cite{Bartol96} & 
$R=$ Data/MC$_{\mbox{no~osc}}$\\[1ex]
\hline
{} &{} &{} &{}\\[-1.5ex]
Upthr. & 857 & 1169 & 0.73\\[1ex]
IU & 157 & 285 & 0.55\\[1ex]
ID+UGS & 262 & 375 & 0.70\\[1ex]
\hline
\end{tabular}\label{tab:macro} 
\caption{MACRO events.}
\end{center}
\end{table}

{\bf \boldmath Upthroughgoing muons ($E_\mu > 1$ GeV)} come from 
interactions in the rock below the detector of $\nm$ with
$\langle E_\nu \rangle \sim 50$ GeV. The MC uncertainties arising from the 
neutrino flux, cross section and muon propagation on the
expected flux of muons are estimated to be $\sim 17\%$; 
 this systematic error on the upthroughgoing muons flux is 
mainly a scale uncertainty.

\begin{figure}
 \begin{center}
\mbox{\epsfig{figure=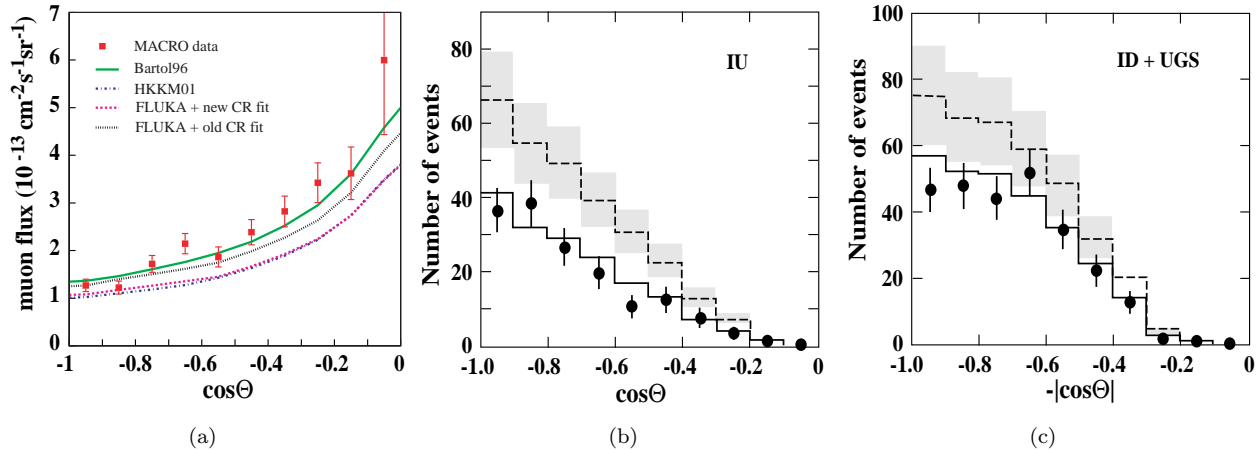,height=5.5cm}}
{\footnotesize (a) \hspace {5cm} (b) \hspace {5cm} (c)}
\caption {(a) Comparison between the zenith distribution of the MACRO 
upthroughgoing muons and the oscillated MC predictions given by
Bartol96 (solid curve), HKKM01 (dash-dotted line), FLUKA fitted to the new CR 
measurements (dashed curve) and FLUKA with the old CR fit  
(dotted curve). Zenith distributions for (b) IU and for (c) ID+UGS MACRO events
(black points) compared with the no oscillation Bartol96 MC (dashed line 
with a scale error band) and with the $\nmnt$ predictions with $\Dm=2.3 \cdot 
10^{-3}$ eV$^2$ and maximal mixing.}
\label{fig:cosze}
\end{center}
\end{figure}

In order to verify that different flux simulations affect the zenith 
distribution at the level of only a few percent (while there is an effect of 
the order of $\sim 25\%$ on the event rates) MACRO compared the predictions
of the Bartol96 \cite{Bartol96}, FLUKA \cite{FLUKA} and 
HKKM01 \cite{honda01} MCs.
 In Fig. \ref{fig:cosze}a the MACRO data are compared with the oscillated
Bartol96, the new HKKM01 and FLUKA calculations using the new CR fit. All
predicted curves are for maximal mixing and $\Dm = 0.0023$ eV$^2$.

{\bf Low energy events.} {\it Semicontained upgoing muons} (IU) come from
$\nm$ interactions inside the lower apparatus. The average 
parent neutrino energy for these events is $\sim 2-3$ GeV. 
 {\it Up stopping muons} (UGS)
are due to external $\nm$ interactions yielding upgoing muons stopping in the 
detector; the {\it semicontained downgoing muons} (ID) are due to downgoing 
$\nm$'s with interaction vertices in the lower detector. The lack
of time information prevents to distinguish between the two subsamples.
An almost equal number of UGS and ID events is expected. The average 
parent neutrino energy for these events is $\sim 2-3$ GeV. 
 The number of events and the angular distributions are compared with the MC
predictions without oscillations in Table \ref{tab:macro} and 
Fig. \ref{fig:cosze}b,c. The low
energy data show a uniform deficit of the measured number of events
over the whole angular distribution with respect to the Bartol96 predictions.

{\bf \boldmath $\nmnt$ against $\nmns$.} Matter effects due to 
the difference between the weak interaction effective potential for 
muon neutrinos with respect to sterile neutrinos, which have null
potential, yield different total number and different zenith 
distributions of upthroughgoing muons. The measured 
ratio between the events with $-1 < \cos \Theta < -0.7$ and with 
$-0.4 < \cos \Theta < 0$ was used \cite{macro-nu}. In this ratio most 
of the theoretical uncertainties
on neutrino flux and cross sections cancel. Combining the experimental
and theoretical errors in quadrature, a global uncertainty of 6\% is obtained.
 The measured ratio is $R_{meas}=1.38$, to be compared with $R_\tau=1.61$ and
$R_{sterile}=2.03$. One concludes that $\nmns$ oscillations (with any mixing) 
are excluded at the 99.8\% C.L. compared to the $\nmnt$ 
channel with maximal mixing and $\Dm=2.3 \cdot 10^{-3} $ eV$^2$.

{\bf \boldmath $\nm$ energy estimate by Multiple Coulomb Scattering of muons.}
Since MACRO was not equipped with a magnet, 
 the only way to estimate the muon energy is through their Multiple Coulomb 
Scattering (MCS) in the absorbers. Two analyses were 
performed \cite{macro-mulsca}. The first was made studying the deflection of 
upthroughgoing muons using the streamer tubes in digital mode. This method 
had a spatial resolution of $\sim 1$ cm. The second analysis was performed 
using the streamer tubes in ``drift 
mode" \cite{macro-mulsca}. The space resolution was $\simeq 3$ mm. For 
each muon, 7 MCS variables were defined and given in input to a Neural 
Network, previously trained with MC events of known energy 
crossing the detector at different zenith angles. The output of this program
gave the muon energy estimate event by event. The sample of upthroughgoing 
muons was separated in 4 subsamples with 
average energies $E_\mu$ of 12, 20, 50 and 100 GeV. The ratios 
Data/MC$_{\mbox{no~osc}}$ as a function of 
$\log_{10} (L/E_\nu)$ obtained from upthroughgoing muons are plotted in 
Fig. \ref{fig:le-contour}a; they are in agreement with the 
$\nmnt$ oscillation hypothesis \cite{macro-nu}.

{\bf New determination of the oscillation parameters.} In previous 
analyses MACRO
fitted the shape of the upthroughgoing muon zenith distribution and the 
absolute flux compared to Bartol96. This yielded $\Dm=2.5 \cdot 10^{-3}$ 
eV$^2$ and maximal mixing \cite{macro-nu}. Later, in order to reduce 
the effects of systematic uncertainties in the MC simulations, MACRO 
 used the following three independent ratios. It was checked
 that FLUKA, HKKM01 and Bartol96 Monte Carlo simulations yield the same 
predictions to within $\sim 5 \%$.

        (i) High Energy Data: zenith distribution ratio: 
$R_1 = N_{vert}/N_{hor}$
        (ii) High Energy Data, $\nu$ energy measurement ratio: 
$R_2 = N_{low}/N_{high}$
        (iii) Low Energy Data:  
$R_3 = (Data/MC)_{IU}/(Data/MC)_{ID+UGS}$.

\noindent The no oscillation hypothesis has a probability 
$P \sim 3 \cdot 10^{-7} $
and is thus ruled out by $ \sim 5 \sigma$. By fitting the 3 ratios 
to the $\nmnt$ oscillation formulae, MACRO obtained $\s2t = 1,~\Dm = 2.3
\cdot 10^{-3}$ eV$^2$ and the allowed region indicated by the solid line 
in Fig. \ref{fig:le-contour}b. There is a good consistency between the old and 
new methods. 

Using the Bartol96 flux, it is possible to add the 
information on the absolute flux values of the

   (iv) High energy data (systematic error $\simeq 17 \%$) 
$R_4 = N_{meas}/N_{MC}$.
   (v) Low energy semicontained muons (scale error $21 \%$) 
$R_5 = N_{meas}/N_{MC}$.

\noindent These informations reduce the area of the allowed region in the 
$ \s2t - \Dm $ plane, as indicated by the dashed line in Fig. 
\ref{fig:le-contour}b. The final MACRO best fit is $\Dm= 2.3 \cdot 10^{-3}$ 
eV$^2$ and $\s2t =1$ ($6\sigma$ significance).

\section{Results from the SuperKamiokande experiment}\label{sec:SK}
SuperKamiokande \cite{skam-new} is a large cylindrical water 
Cherenkov detector of 39 m diameter and 41 m height containing 50 kt of 
water (the fiducial mass of the detector for atmospheric neutrino analyses 
is 22.5 kt); it was seen by 11146, 50-cm-diameter 
inner-facing phototubes. The 2 m thick outer layer of water, acting 
as an anticoincidence, was seen by 1885 smaller outward-facing
photomultipliers. The ultra pure water
has a light attenuation of almost 100 m. The detector 
is located in the Kamioka mine, Japan, under 2700 m.w.e. SK took data in 
its full configuration from April 1996 till November 2001, when an accident 
happened. It resumed operation with about half of PMTs in October 2002.

\begin{figure}
 \begin{center}
\mbox{\epsfig{figure=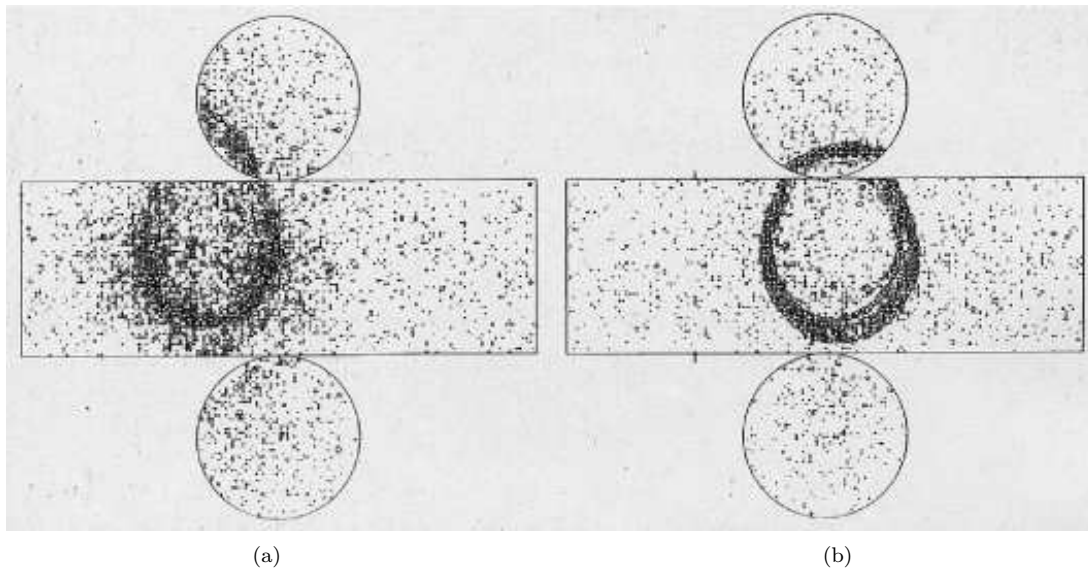,height=7cm}}
{\footnotesize \hspace{5cm} (a) \hspace{7cm} (b)}
\caption {Sharpness of Cherenkov rings produced in the SK detector by (a)
an electron and (b) a muon.}
\label{fig:sk_topo}
\end{center}
\end{figure}

Atmospheric neutrinos are detected in SK by measuring the Cherenkov light
generated by the charged particles produced in the neutrino
CC interactions with the protons and oxygen nuclei.
 The large detector mass and the possibility of clearly
defining a large inner volume allow to collect a high
statistics sample of {\it fully contained} events (FC) up to relatively high
energies (up to $\sim 5$ GeV). The FC events have both the
neutrino vertex and the resulting particle tracks entirely within the
fiducial volume; they yield rings of Cherenkov light on the PMTs.
 Fully contained events can be further subdivided
into two subsets, the so-called {\it sub-GeV} and {\it multi-GeV} events,
with energies below and above 1.33 GeV, respectively.

\begin{table}
\begin{center}
\begin{tabular}{@{}ccc@{}}
\hline
{} &{} &{}\\[-1.5ex]
{} &Data & MC$_{\mbox{no~osc}}$ \cite{honda01}\\[1ex]
\hline
{} &{} &{}\\[-1.5ex]
Sub-GeV 1-ring & 3227 & 4213\\[1ex]
Sub-GeV multi-ring &  208 & 323\\[1ex]
Multi-GeV 1-ring & 651 & 900\\[1ex]
Multi-GeV multi-ring & 439 & 712\\[1ex]
PC $\mu$ & 647 & 1034\\[1ex]
Up-stop & 418 & 721\\[1ex]
Upthr. & 1842 & 1684\\[1ex]
\hline
\end{tabular}\label{tab:superk} 
\caption{SK $\mu$ data taken from 1996 to 2001.}
\end{center}
\end{table}

Another sub-sample, defined as the {\it partially contained} events (PC), is
represented by those CC interactions where the vertex is still within the 
fiducial volume, but at least a primary charged particle, typically
the muon, exits the detector without releasing all of its energy. In this case
the light pattern is a filled circle.
 For these events the energy resolution is worse than
for FC interactions. {\it Upward-going muons} (UPMU), produced by 
neutrinos coming from
below and interacting in the rock, are further subdivided into 
{\it stopping  muons} ($\langle E_\nu \rangle \sim 7$ GeV) and 
{\it upthroughgoing muons} ($\langle E_\nu \rangle \sim 70 \div 80$ GeV), 
 according to whether or not they stop in the detector. 

\begin{figure}[ht]
 \begin{center}
\mbox{\epsfig{figure=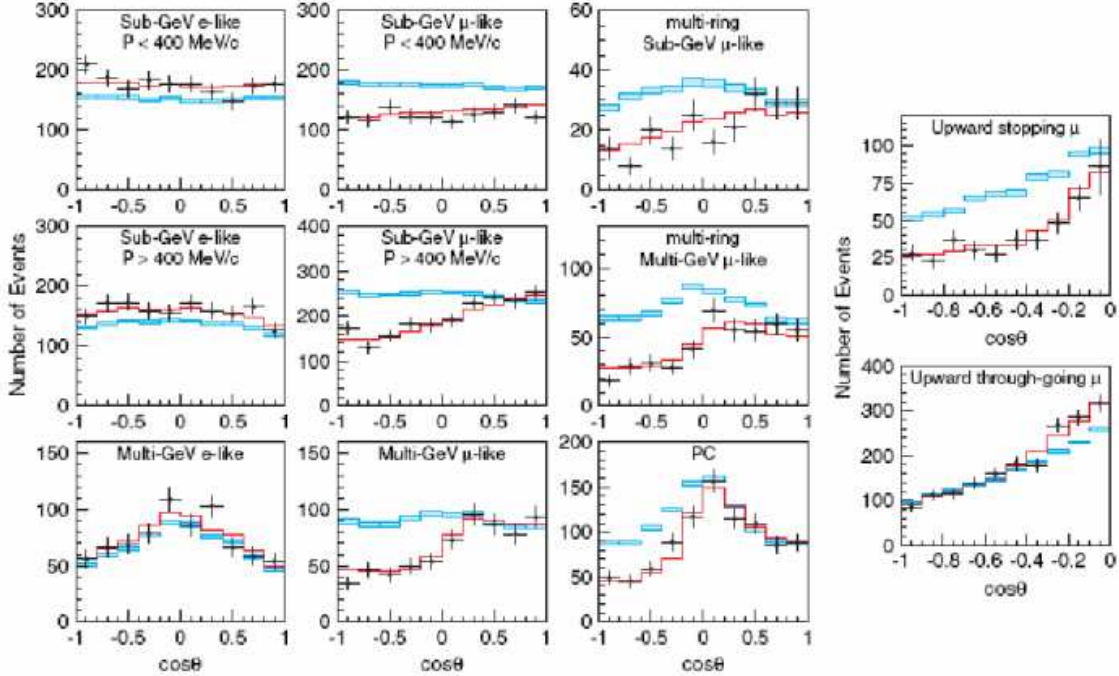,height=10cm}}
\caption {SK data taken from 1996 till 2001 with the detector in full
configuration (1489 days for FC+PC events and 1646 days for 
upgoing $\mu$). Zenith distributions for
 $e$-like and $\mu$-like sub-GeV and multi-GeV 
events, for partially contained events and for upthroughgoing and 
stopping muons (black points). The boxes are the no oscillation HKKM01 
predictions, the solid lines refer to $\nmnt$ oscillations with maximal 
mixing and $\Dm=2.4\cdot 10^{-3}$ eV$^2$.}
\label{fig:sk_zenith}
\end{center}
\end{figure}

Particle identification in SuperKamiokande is performed using
likelihood functions to parametrise the sharpness of
the Cherenkov rings, which are more diffuse for electrons than for
muons, Fig. \ref{fig:sk_topo}. The algorithms are able to discriminate
the two flavours with high purity
(of the order of $98\%$ for single track events). 
The zenith angle distributions for $e$-like and $\mu$-like 
sub-GeV and multi-GeV events, for PC events and for upward 
throughgoing or stopping muons are shown in Fig. \ref{fig:sk_zenith}. These
data were taken from 1996 till 2001 with the detector in full
configuration (1489 days for FC+PC events and 1646 days for 
upgoing $\mu$). The data and MC behaviour shows the problem with 
the new HKKM01 MC discussed in Sec. \ref{sec:MC}. The 
number of measured and 
expected $\mu$-like events are summarised in Table \ref{tab:superk}. 
 The new data, taken in 2003 with about half of the PMTs and referring to 
311 days for FC+PC events and 243 days for upgoing muons, show the same 
behaviour \cite{skam-new} as the older data.

\begin{figure}
 \begin{center}
\mbox{\epsfig{figure=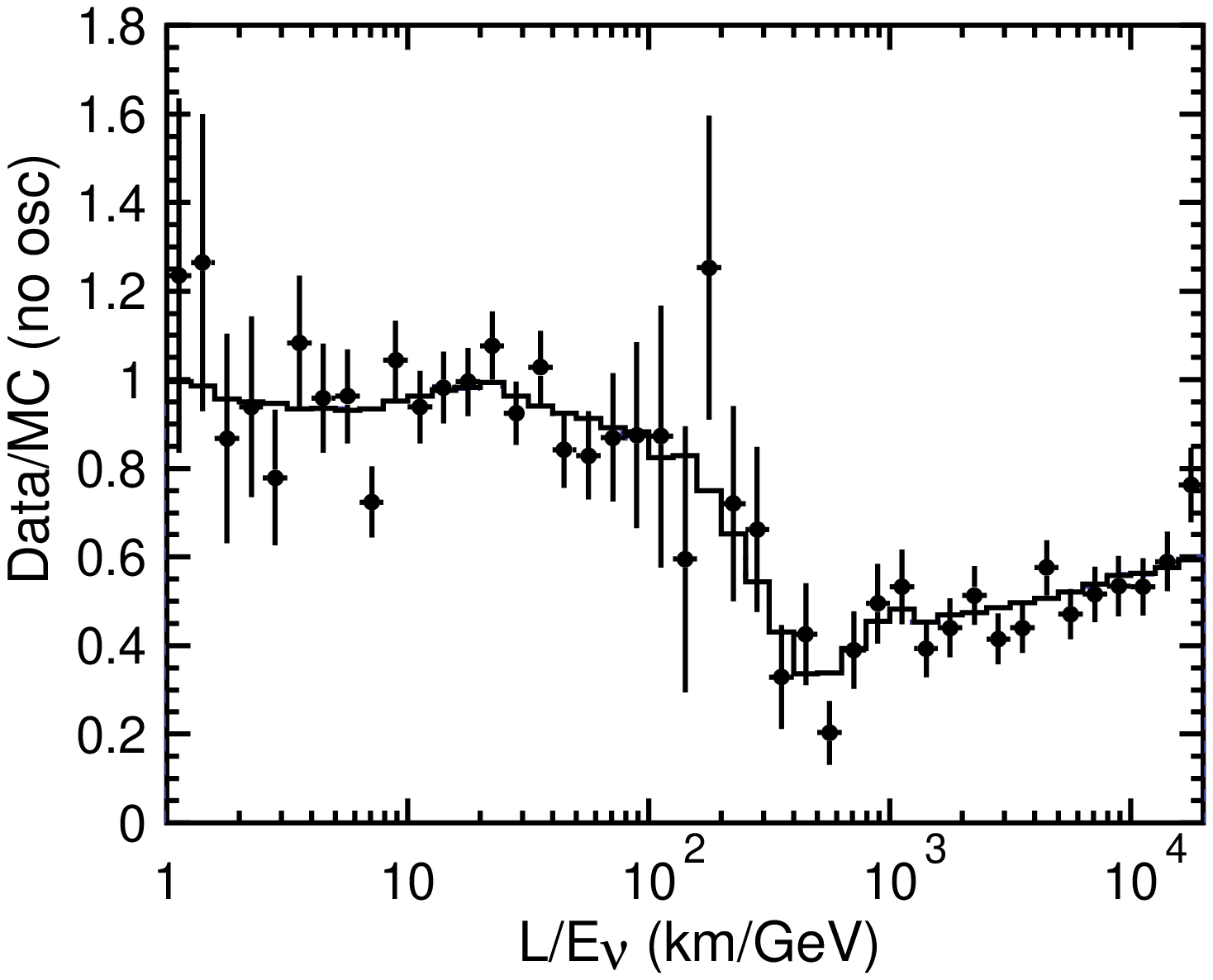,height=6cm}}
        \hspace{2mm}
\mbox{\epsfig{figure=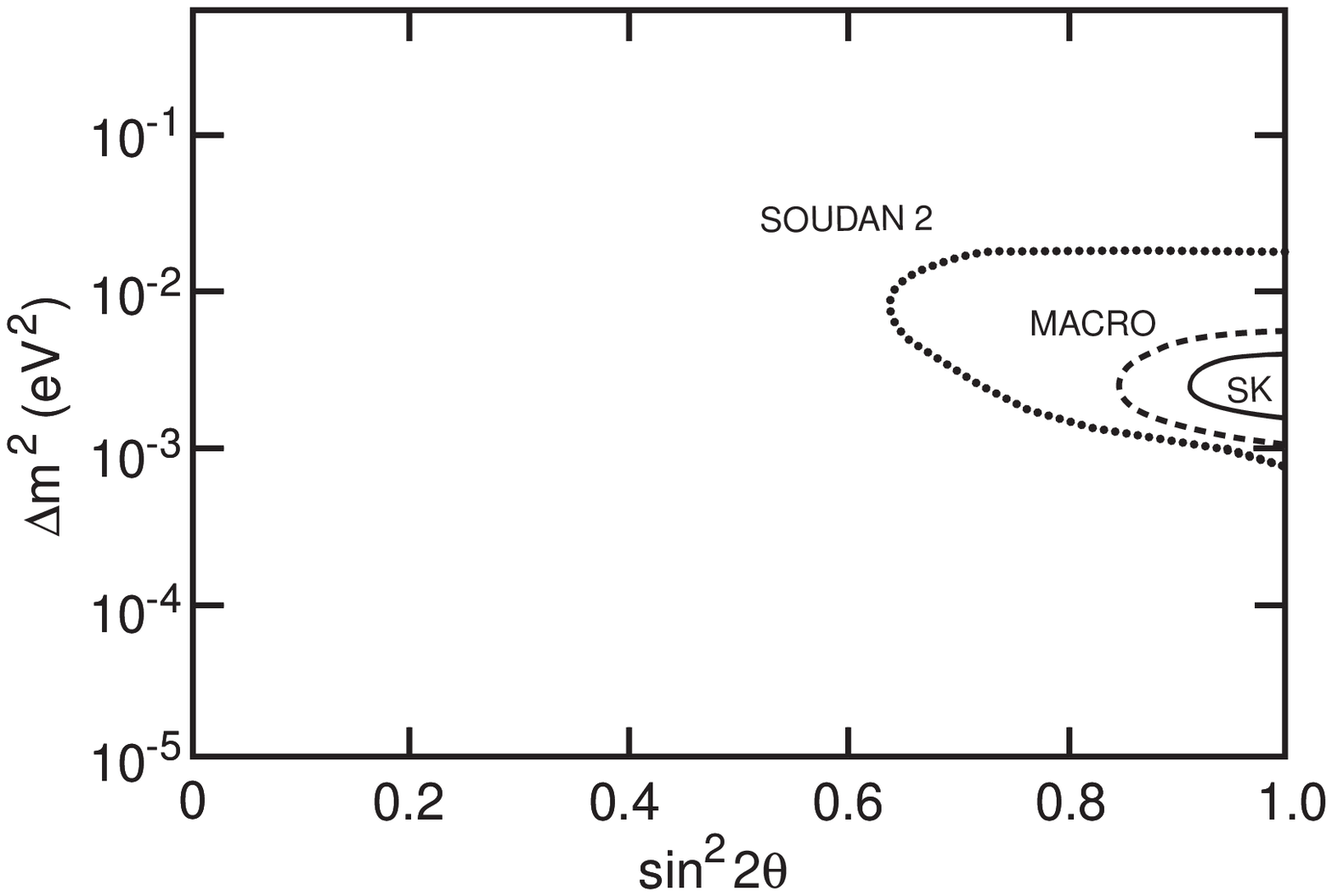,height=5.8cm}}
{\footnotesize (a) \hspace{8cm} (b)}
\caption{(a) Ratio of the data to the MC events without neutrino oscillation 
(black points) as a function of the reconstructed $L/E_\nu$ compared with the 
best-fit expectation for 2-flavour $\nmnt$ oscillations (solid line). 
 The error bars are statistical only. (b) 90\% C.L. allowed regions for $\nmnt$
oscillations obtained by the SK, MACRO and Soudan 2 experiments.}
\label{fig:sk_le_contour}
\end{center}
\end{figure}

New analyses have been performed leaving free the normalisation.
The last value for the double ratio $R^\prime$ reported by SK is
$0.658 \pm 0.016_{stat} \pm 0.032_{sys}$ for the sub-GeV sample and
$0.702 \pm 0.031_{stat} \pm 0.099_{sys}$ for the multi-GeV sample
(both FC and PC).

SK used also a selected sample of events with good resolution in
$L/E_\nu$, to search for the dip in the oscillation probability expected 
when the argument of the second sine-squared term in Eq.~(\ref{eq:prob_osc})
is $\pi/2$. A dip in the $L/E_\nu$ distribution is observed at
$L/E_\nu \simeq 500$ km/GeV, see Fig. \ref{fig:sk_le_contour}a. This 
is another proof in favour of $\nu$ oscillations and a further constraint 
on $\Dm$. Alternative 
models that could explain the zenith angle and energy dependent deficit 
of the atmospheric muon neutrinos are disfavoured, since they do not predict 
any dip in the $L/E_\nu$ distribution \cite{skam-new}. 

 Interpreting the $\mu$-like event deficit as the result of $\nmnt$
oscillations in the two-flavour mixing scheme, SK computed
an allowed domain for the oscillation parameters \cite{skam-new}, see 
Fig. \ref{fig:sk_le_contour}b. The events were binned in a multi-dimensional 
space defined by particle
type, energy and zenith angle, plus a set of parameters to account for
systematic uncertainties. The best fit using FC, PC, UPMU and MRING 
events \cite{skam-new} corresponds to maximal mixing and $\Dm =
2.4 \cdot 10^{-3}$ eV$^2$. In Fig. \ref{fig:sk_le_contour}b, the 90\% SK 
allowed region in the $\s2t - \Dm$ plane is compared with the MACRO 
and Soudan 2 ones. The limit lines represent smoothed interpolations 
and are qualitative.
 
\section{Conclusions}\label{sec:conclu}
The atmospheric neutrino data strongly favour $\nmnt$ oscillations with 
maximal mixing and $\Dm =0.0023 - 0.0052$ eV$^2$ (Soudan 2: 0.0052, 
 MACRO: 0.0023, SK: 0.0024 eV$^2$).  MACRO and SK exclude $\nmns$ 
oscillations at the level of 99.9\%; SK excludes also $\nmne$ oscillations
and some exotic processes \cite{skam-new,exotic}. The K2K \cite{k2k} 
long baseline 
experiment, using the $\nm$ beam produced at KEK and detected by 
SK, supports the $\nmnt$ oscillations with $\Dm =0.0027$ eV$^2$. SK 
finds the predicted dip in the $L/E_\nu$ distribution. 

It has been hypothesized that, besides the dominant mass neutrino oscillations,
there could be sub-dominant oscillations due to possible Lorentz 
Invariance Violation (LIV) \cite{lorentz} (or violation of the equivalence 
principle). In this case, one would consider a mixing of 
flavour eigenstates and velocity eigenstates and estimate upper limits on
the LIV parameters $\delta v/2=(v_3-v_2)/2$ and $\s2t $$_v$. Preliminary 
analyses of SK and MACRO high energy muon data yield upper
limits for $\delta v/2$ at the level
of $\sim 10^{-23}$ for low mixing angles and $\sim 10^{-26}$ for large 
mixing angles \cite{deltav}. 

Several other possibilities have been discussed, also possible $\nu$
decays \cite{deltav,notte}.

\section{Acknowledgements}
We would like to acknowledge the cooperation of many experimental and 
theoretical colleagues from the MACRO, Soudan 2 and SK experiments. Particular
thanks to the colleagues of the Bologna group.

\end{document}